\documentclass[12pt]{article}
\usepackage{amsmath}
\usepackage{graphicx}
\usepackage{enumerate}
\usepackage{natbib}
\usepackage{url} 
\usepackage{ulem}
\usepackage{booktabs}
\usepackage{adjustbox}
\usepackage{multirow}

\newcommand{\biblist}{\begin{list}{}
{\listparindent 0.0cm \leftmargin 0.50cm \itemindent -0.50 cm
\labelwidth 0 cm \labelsep 0.50 cm
\usecounter{list}}\clubpenalty4000\widowpenalty4000}
\newcommand{\ebiblist}{\end{list}}

\newtheorem{thm}{Theorem}[section]

\newcommand{\E}{\operatorname{E}}
\newcommand{\V}{\operatorname{var}}
\newcommand{\T}{\operatorname{T}}

\newcommand{\N}{\operatorname{N}}
\newcommand{\OP}{\operatorname}

\title{Enhanced Laplace Approximation} 
\author{Jeongseop Han \and Youngjo Lee}
\date{} 
\begin{document}

\baselineskip .3in
\maketitle 

\begin{abstract}
The Laplace approximation (LA) has been proposed as a method for approximating the marginal likelihood of statistical models with latent variables. However, the approximate maximum likelihood estimators (MLEs) based on the LA are often biased for binary or spatial data, and the corresponding Hessian matrix underestimates the standard errors of these approximate MLEs. A higher-order approximation has been proposed; however, it cannot be applied to complicated models such as correlated random effects models and does not provide consistent variance estimators. In this paper, we propose an enhanced LA (ELA) that provides the true MLE and its consistent variance estimator. We study its relationship to the variational Bayes method. We also introduce a new restricted maximum likelihood estimator (REMLE) for estimating dispersion parameters. The results of numerical studies show that the ELA provides a satisfactory MLE and REMLE, as well as their variance estimators for fixed parameters. The MLE and REMLE can be viewed as posterior mode and marginal posterior mode under flat priors, respectively. Some comparisons are also made with Bayesian procedures under different priors.
\end{abstract}

\baselineskip .3in

\baselineskip .3in

\baselineskip .3in

\baselineskip .3in

\baselineskip .3in

\baselineskip .3in

\baselineskip .3in

\baselineskip .3in

\newpage

\section{Introduction}
\label{sec:ELA_intro}

\cite{lee1996} proposed the use of the h-likelihood for making inferences about statistical models with latent variables which are widely used in various fields. Consider a hierarchical generalized linear model (HGLM) with $\OP{E}(y|z)=\mu $, $\OP{var}(y|z)=\phi V(\mu)$, and the linear predictor 
\begin{equation*}
\eta =g(\mu)=X\beta +L(\Sigma)z,
\end{equation*}
where $V(\mu)$ is the variance function, $\beta$ indicates fixed effects, $z$ indicates latent variables, namely random effects, and $\tau =(\phi, \Sigma)$ are dispersion parameters. The h-likelihood of the HGLM is written as
\begin{equation*}
H(\theta, z) = f_{\theta}(y, z)=f_{\theta }(y \mid z)f(z).
\end{equation*}
The h-likelihood consists of three objects: the observed data $y$, fixed unknown parameters $\theta =(\beta, \tau)$, and unobserved latent variables $z$. The marginal likelihood can be used to estimate the fixed parameters $\theta$ by integrating out the latent variables from the h-likelihood:
\begin{equation}
L_{m}(\theta)=f_{\theta}(y)=\int H(\theta, z)dz.  \label{eq:marginal_lik}
\end{equation}
To make inferences about the random effects $z$, \cite{leenelderpawitan2017} proposed the use of the predictive likelihood: 
\begin{equation*}
L_{p}(z|y; \theta) = f_{\theta}(z \mid y)=f_{\theta}(y, z)/f_{\theta}(y) = H(\theta, z)/L_{m}(\theta),
\end{equation*}
which is analogous to the use of a Bayesian posterior under a flat prior on $\theta$.

In random effects models, the h-likelihood can be explicitly written, whereas the marginal and predictive likelihoods often involve intractable integration. The Gauss-Hermite quadrature can be used for the integral shown in (\ref{eq:marginal_lik}). However, this formulation becomes numerically difficult as the dimension of integration increases \citep{hedeker2006}. Instead, in random effects models, \cite{leenelder2001} proposed the use of the Laplace approximation (LA) \citep{tierney1986}, which is widely used and has been implemented by various packages \citep{rue2009, kristensen2016, dhglm2018}. Recently, \cite{perry2017} proposed a fast moment-based method for random effects models, which does not allow correlated random effects and is restricted to nested random effects models. Thus, this method cannot be used for crossed random effects models. In this paper, for the maximum likelihood (ML) estimation, we exploit an alternative expression of the marginal likelihood:
\begin{equation}
\label{eq:marginal_another}
L_{m}(\theta) = H(\theta, z) / L_{p}(z \mid y; \theta).
\end{equation}
For the log-likelihoods we use $h(\theta, z)=\log H(\theta, z)$, $\ell_{m} (\theta)=\log L_{m}(\theta)$, and $\ell_{p}(z|y;\theta )=\log L_{p}(z|y;\theta )$. 

\cite{leenelder2001} extended the restricted likelihood \citep{patterson1971} for normal linear mixed models to HGLMs, which is important for estimating the dispersion parameter $\tau$. However, there is no theoretical justification that the current approximate maximum likelihood estimator (MLE) and restricted maximum likelihood estimator (REMLE), which are based on the LA, are asymptotically equivalent to the true MLE and REMLE. Furthermore, how their consistent variance estimators could be obtained remains ambiguous. In this paper, we propose the use of an integrated likelihood as a new restricted likelihood and introduce the enhanced LA (ELA), which provides the MLE, REMLE, and their consistent variance estimators. 

\section{Review of the LA}
\label{section:ELA_LA} 
Throughout the paper, we impose the following regularity conditions: 
\begin{itemize}
    \item[R1.] The parameter space $\Theta $ is convex.
    \item[R2.] All likelihoods are smooth and unimodal with respect to $\theta $.
\end{itemize}
The LA to the marginal likelihood $L_{m}(\theta)$ is
\begin{equation*}
\hat{L}_{m}(\theta )=H(\theta ,\tilde{z})\left| \frac{1}{2\pi }\tilde{\Omega}_{zz}\right| ^{-\frac{1}{2}},  \label{eq:Lap}
\end{equation*}
where $\tilde{z}=\OP{arg}\max_{z}h(\theta ,z)=\OP{arg}\max_{z}\ell_{p}(z|y;\theta )$ and
\begin{equation*}
\tilde{\Omega}_{zz}=-\frac{\partial ^{2}}{%
\partial z\partial z^{\OP{T}}} h(\theta ,z)\Big |_{z=\tilde{z}}=-\frac{\partial ^{2}}{\partial z\partial z^{\OP{T}}}\ell _{p}(z|y;\theta )\Big |_{z=\tilde{z}}.
\end{equation*}
According to (\ref{eq:marginal_another}), the LA to $L_{m}(\theta)$ can be defined as
\begin{equation*}
\hat{L}_{m}(\theta ) = H(\theta, \tilde{z}) / \hat{L}_{p}(\tilde{z}|y; \theta),
\end{equation*}
This formulation can be viewed as the use of an approximate predictive likelihood $\hat{L}_{p}(z|y; \theta)$ in (\ref{eq:marginal_another}), based on the normal distribution
\begin{equation}
z\mid y \sim \OP{N}\left( \tilde{z},\tilde{\Omega}_{zz}^{-1}\right).  \label{eq:predictive_lik_LA}
\end{equation}
This gives
\begin{equation*}
\hat{\ell}_{m}(\theta )=\log \hat{L}_{m}(\theta )=h(\theta ,\tilde{z})-\hat{\ell}_{p}(\tilde{z}|y;\theta )=h(\theta ,\tilde{z})-\frac{1}{2}\log \left| \frac{1}{2\pi }\tilde{\Omega}_{zz}\right|.  
\end{equation*}
Thus, the LA is exact when the predictive likelihood is normal. 
Let $\hat{\theta}$ be the MLE and $\hat{\theta}^{\OP{L}}$ be the approximate MLE, which are modes of $\ell_{m} (\theta )$ and $\hat{\ell}_{m}(\theta )$, respectively. As the sample size $n \to \infty$, if $\hat{\theta}\overset{\OP{P}}{\rightarrow }\theta _{0}$ and
\begin{equation}
\ell_{m} (\theta )-\hat{\ell}_{m}(\theta )\overset{\OP{P}}{\rightarrow }0,\text{ uniformly in }\theta ,  \label{eq:uniform_lik}
\end{equation}
then $\hat{\theta}^{\OP{L}}\overset{\OP{P}}{\rightarrow }\theta _{0}$.
However, in general, it is difficult to justify that the LA $\hat{\ell}_{m}(\theta)$ satisfies the uniform convergence condition (\ref{eq:uniform_lik}). Let $\theta_{0}^{\OP{L}}$ be the probability limit of $\hat{\theta}^{\OP{L}}$. If $\sqrt{n} ( \hat{\theta}^{\OP{L}} - \theta_{0}^{\OP{L}} ) =O_{p}(1)$, then
\begin{equation}
\sqrt{n}\left( \hat{\theta}^{\OP{L}}-\theta_{0}^{\OP{L}}\right) \overset{%
\OP{d}}{\rightarrow } \OP{N}\left\{ 0,\mathcal{G}^{-1}\left( \theta_{0}^{\OP{L}} \right) \right\} ,  \label{eq:var_Godambe}
\end{equation}
where $\tilde{\mathcal{G}}(\theta )=\tilde{\mathcal{H}}(\theta )\tilde{\mathcal{K}}%
^{-1}(\theta )\tilde{\mathcal{H}}(\theta )$, $\tilde{\mathcal{H}}(\theta )=%
\OP{E} \{ -\partial^{2} \hat{\ell}_{m}(\theta ) / \partial \theta
\partial \theta^{\OP{T}} \} $, $\tilde{\mathcal{K}}(\theta )=\OP{var%
} \{ \partial \hat{\ell}_{m}(\theta ) / \partial \theta  \} $ and $\mathcal{G}( \theta_{0}^{\OP{L}}) =\lim_{n\rightarrow \infty }n^{-1}\tilde{\mathcal{G}}( \theta_{0}^{\OP{L}}) $. \cite{kristensen2016} and \cite{leenelderpawitan2017} proposed the use of the inverse Hessian matrix of $\hat{\ell}_{m}(\theta )$ as a variance estimator of $\hat{\theta}^{\OP{L}}$. \cite{ogden2017} provided regularity conditions that guarantee asymptotic equivalence between the Hessian matrix of $\hat{\ell}_{m}(\theta )$ and that of $\ell_{m} (\theta )$. However, these conditions are hardly satisfied. As mentioned in \cite{bologa2021}, the Bayesian approach views the approximate MLE $\hat{\theta}^{\OP{L}}$ as an approximate mode of the posterior distribution under a flat prior on $\theta$. \cite{pauli2011} further showed that
\begin{equation*}
\sqrt{n}\left( \theta -\hat{\theta}^{\OP{L}}\right) \mid y\overset{\OP{d}}{\rightarrow }\OP{N}\left\{ 0,\mathcal{H}^{-1}\left( \theta_{0}^{\OP{L}} \right) \right\}, 
\end{equation*}
where 
\begin{equation*}
\mathcal{H}\left( \theta_{0}^{\OP{L}} \right) =\lim_{n\rightarrow \infty }%
\frac{1}{n}\left\{ -\frac{\partial ^{2}}{\partial \theta \partial \theta ^{%
\OP{T}}}\hat{\ell}_{m}(\theta )\Big |_{\theta =  \theta_{0}^{\OP{L}} }\right\} .
\end{equation*}
Thus, the variance estimators presented by \cite{kristensen2016} and \cite{leenelderpawitan2017} can be viewed as estimating the variance of the approximate Bayesian posterior mode $\hat{\theta}^{\OP{L}}$; see the numerical study of \cite{bologa2021}. In addition, \cite{jinlee2022} investigated the frequentist sandwich variance estimator (\ref{eq:var_Godambe}) of the approximate MLE $\hat{\theta}^{\OP{L}}$.

Assume that $d$ is the dimension of the integral in (\ref{eq:marginal_lik}). The LA is valid in the sense that $\ell_{m} (\theta )-\hat{\ell}_{m}(\theta )=o_{p}(1)$ when $d=o(n^{1/3})$ \citep{shunmc1995, ogden2021}; thus, the LA may not be suitable for crossed random effects models with $d=O(n^{1/2})$ and correlated random effects models with $d=O(n)$. Furthermore, the performance of the LA is often unsatisfactory for binary outcomes \citep{shun1997}. Thus, \cite{shunmc1995} proposed the use of the second-order LA in the exchangeable binary array model for salamander mating data. \cite{shun1997} investigated parameter estimation based on the second-order LA. However, due to the complexity of the approximation, the author could compute only some selected terms. \cite{nohlee2007} showed how to compute all the terms in the second-order LA and developed a REML estimation procedure for salamander mating data. However, the second-order LA can be applied to a limited class of models due to the complexity of the approximation. Furthermore, even if the second-order LA is applicable, the approximation is often slow because a considerable number of terms must be computed. 

In summary, (i) $\ell_{m} (\theta ) - \hat{\ell}_{m}(\theta ) \neq o_{p}(1)$ as $d$ increases, and (ii) even if $\ell_{m} (\theta ) - \hat{\ell}_{m}(\theta ) = o_{p}(1)$, the approximate MLE $\hat{\theta}^{\OP{L}}$ may not be the MLE $\hat{\theta}$. Furthermore, (iii) it is not known how to obtain a consistent variance estimator for $\hat{\theta}$. (iv) It is also of interest to have REMLEs for dispersion parameters. A general higher-order LA may not be sufficient for resolving these problems.

\section{ELA}
\label{section:ELA_ELA} 
Assume that $q(z)$ is an arbitrary density function with $\int q(z)dz=1$ that has the same support as the predictive likelihood $L_{p}(z | y; \theta)$. Next, from (\ref{eq:marginal_lik}) the marginal likelihood is defined as
\begin{equation*}
    L_{m}(\theta) = \int H(\theta, z) dz = \int \frac{H(\theta, z)}{q(z)} q(z) dz.
\end{equation*}
Thus, we can approximate the marginal likelihood as
\begin{equation*}
\tilde{L}_{B}(\theta )=\frac{1}{B}\sum_{b=1}^{B}\frac{H(\theta ,Z_{b})}{q(Z_{b})},  
\end{equation*}
where $Z_{b}$ are iid samples from $q(z)$. Since $H(\theta ,Z_{b})/q(Z_{b})$ can be viewed as iid samples with the mean $L_{m}(\theta )$, $\tilde{L}_{B}(\theta )$ is a consistent estimator of $L_{m}(\theta)$, i.e., as $B \to \infty$,
\begin{equation*}
    \tilde{L}_{B}(\theta )\overset{\OP{P}}{\to } L_{m} (\theta ).
\end{equation*}

The variational Bayes method has been proposed for approximating $\ell_{m}(\theta)$ \citep{kingma2013}. For any $q(z)$, 
\begin{eqnarray*}
\ell_{m} (\theta ) &=& \int \log \left\{ \frac{H(\theta, z)}{q(z)}\right\} q(z)dz + R \nonumber \\
&\geq& \int \log \left\{ \frac{H(\theta, z)}{q(z)}\right\} q(z)dz = \ell_{v}(\theta; q), 
\end{eqnarray*}
where
\begin{equation*}
    R = \int \log \left\{ \frac{q(z)}{L_{p}(z \mid y; \theta)}\right\} q(z)dz \geq 0,
\end{equation*}
and $\ell_{v}(\theta; q)$ is referred to as the evidence lower bound (ELBO). The marginal log-likelihood in (\ref{eq:marginal_lik}) can be approximated by maximizing the ELBO
\begin{equation*}
    \hat{\ell}_{v} (\theta) = \max_{q} \ell_{v}(\theta; q).
\end{equation*}
In the variational Bayes methods, $q(z)$ is often assumed to have a normal density $\OP{N}(\mu ,\Gamma )$ with an arbitrary mean $\mu $ and arbitrary covariance matrix $\Gamma $. In general, the ELBO is not a tight lower bound, i.e., $\ell_{m}(\theta) - \hat{\ell}_{v} (\theta) > 0$ since $R>0$. To address this issue, \cite{burda2016} modified the ELBO as follows:
\begin{equation*}
    \tilde{\ell}_{v, B}(\theta; \mu, \Gamma) = \E_{\mu, \Gamma} \left \{ \log \tilde{L}_{B}(\theta) \right \},
\end{equation*}
where $Z_{b}$ are iid samples from $\N(\mu, \Gamma)$. The authors used the seesaw algorithm: (i) given $\theta$, update ($\mu, \Gamma$) by maximizing $\tilde{\ell}_{v, B}(\theta; \mu, \Gamma)$ and (ii) given ($\mu, \Gamma$), update $\theta$ by maximizing $\tilde{\ell}_{v, B}(\theta; \mu, \Gamma)$. In correlated random effects models with $d=n$, estimating $\mu$ and $\Gamma$ is not straightforward. The ELBO has been studied to approximate the marginal log-likelihood. However, the main interest of this paper is how to obtain the true MLE $\hat{\theta}$ and its consistent variance estimator in general cases.

According to the expression (\ref{eq:marginal_another}), if the value of $L_{p}(z^{\ast }|y;\theta )$ is known at any
point $z^{\ast }$, it is immediate that $L_{m}(\theta )=H(\theta ,z^{\ast
})/L_{p}(z^{\ast }|y;\theta )$. However, in general, $L_{p}(z|y;\theta )$ is not
known for all $z$. Recall that the LA approximates the predictive likelihood $L_{p}(z|y; \theta)$ at $\tilde{z}$ by $\hat{L}_{p}(\tilde{z}|y;\theta )$ as
\begin{equation*}
\hat{L}_{m}(\theta )=H(\theta ,\tilde{z})/\hat{L}_{p}(\tilde{z}|y;\theta ).
\end{equation*}
Since $\ell_{m}(\theta) - \hat{\ell}_{m}(\theta) = \hat{\ell}_{p}(\tilde{z}|y; \theta) - \ell_{p}(\tilde{z}|y; \theta)$, the accuracy of the LA is the same as that of the predictive likelihood $\hat{L}_{p}(z | y; \theta)$. Let
\begin{equation*}
\hat{L}_{B}(\theta )=\frac{1}{B}\sum_{b=1}^{B}\hat{L}_{m}(\theta ;Z_{b}),
\end{equation*}
where $\{Z_{b}:b=1,\ldots ,B\}$ are iid samples from $\OP{N}( \tilde{z}, \tilde{\Omega}_{zz}^{-1})$ and
\begin{equation*}
\hat{L}_{m}(\theta ;Z)=H(\theta ,Z)/\hat{L}_{p}(Z|y;\theta ).
\end{equation*}
The LA is $\hat{L}_{B}(\theta )$ with $B=1$ at $Z_{b}=\tilde{z}$. We call $\hat{L}_{B}(\theta )$ the \textit{ELA} when $q(z)$ is the density function of $\N(\tilde{z}, \tilde{\Omega}_{zz}^{-1})$. In the Appendix, we show that if the true predictive likelihood $L_{p}(z|y;\theta )$ is normal, then, for all $B\geq 1$
\begin{equation}
\label{eq:normal_equiv}
    \hat{L}_{B}(\theta)=L_{m}(\theta).
\end{equation}
If $\hat{L}_{p}(z|y; \theta)$ is close to the true $L_{p}(z|y; \theta)$, we expect that $\hat{L}_{B}(\theta)$ provides an accurate estimate of $L_{m}(\theta)$ for small values of $B$. As the LA provides an accurate approximation of $L_{m}(\theta)$, the use of $\OP{N}(\tilde{z}, \tilde{\Omega}_{zz}^{-1})$ as $q(z)$ is preferred. \cite{burda2016} improved the variational method by exploiting the expression (\ref{eq:marginal_lik}) of the marginal likelihood. The ELA further improves the variational method by using the alternative expression (\ref{eq:marginal_another}).

\begin{thm}
\label{thm:ELA}
Let $\hat{\ell}_{B}(\theta) = \log \hat{L}_{B}(\theta)$ and $\hat{\theta}_{B}^{\OP{ELA}}=\OP{arg}\max_{\theta} \hat{\ell}_{B}(\theta )$. Under regularity conditions R1 and R2, as $B \rightarrow \infty$,
\begin{equation*} \hat{\theta}_{B}^{\OP{ELA}}\overset{\OP{P}}{\to }\hat{\theta}.
\end{equation*}
\end{thm}

Now, we study how to obtain a consistent estimator for the information matrix
\begin{equation*}
    I(\theta) = - \frac{\partial^{2} \ell_{m}(\theta)}{\partial \theta \partial \theta^{\T}}.
\end{equation*}
Let $\hat{I}_{B} = I_{B}(\hat{\theta}_{B}^{\OP{ELA}})$, where
\begin{eqnarray*}
    I_{B}(\theta) &=& \left [ \sum_{b=1}^{B} \left \{ w(\theta, Z_{b}) \frac{\partial h(\theta, Z_{b})}{\partial \theta} \right \} \right ] \left [ \sum_{b=1}^{B} \left \{ w(\theta, Z_{b}) \left ( \frac{\partial h(\theta, Z_{b})}{\partial \theta} \right )^{\T} \right \} \right ] \nonumber \\
    && - \sum_{b=1}^{B} \left [ w(\theta, Z_{b}) \left \{ \frac{\partial h(\theta, Z_{b})}{\partial \theta} \left ( \frac{\partial h(\theta, Z_{b})}{\partial \theta} \right )^{\T} + \frac{\partial^{2} h(\theta, Z_{b})}{\partial \theta \partial \theta^{\T}} \right \} \right ]
\end{eqnarray*}
and $w(\theta, Z_{b}) = \hat{L}_{m}(\theta, Z_{b}) / \sum_{t=1}^{B} \hat{L}_{m}(\theta, Z_{t})$. Then, we have the following theorem.
\begin{thm}
\label{thm:variance}
  As $B \to \infty$, $\hat{I}_{B} \overset{\OP{P}}{\to} I(\hat{\theta})$.
\end{thm}
According to Theorem \ref{thm:variance}, the variance of the MLE $\hat{\theta}$ can be consistently estimated by
\begin{equation*}
    \widehat{\V} \left ( \hat{\theta} \right ) = \hat{I}_{B}^{-1}.
\end{equation*}

\section{Restricted Likelihood}
\label{section:ELA_REML}
For cases in which $\tau$ and $\beta$ are orthogonal, \cite{cox87} proposed the use of an adjusted profile likelihood for the dispersion parameters $\tau$ based on the marginal likelihood $L_{m}(\theta)$:
\begin{equation*}
\hat{R}(\tau )=L_{m} \left( \tau, \tilde{\beta} \right) \left| \frac{1}{2\pi }\tilde{\Omega}_{\beta \beta }\right|^{-\frac{1}{2}} ,
\end{equation*}
where $\tilde{\beta} = \tilde{\beta}(\tau) = \OP{arg} \max_{\beta}L_{m}(\beta, \tau)$ and $\tilde{\Omega}_{\beta \beta }=\{-\partial^{2} \ell_{m} (\beta ,\tau )/\partial \beta \partial \beta ^{\OP{T}}\}|_{\beta =\tilde{\beta}}$. \cite{barndorff-nielsen1987} noted that the Cox-Reid adjusted profile likelihood is the LA to the integrated likelihood 
\begin{equation*}
    R(\tau) = \int L_{m}(\tau, \beta) d \beta = \hat{R}(\tau) (1 + O_{p}(n^{-1})).
\end{equation*}
Under the flat conditional prior $\pi(\beta | \tau) = 1$, \cite{sweeting1987} noted that the integrated likelihood becomes the marginal posterior density of $\tau$:
\begin{equation*}
    R(\tau) = \int L_{m}(\tau, \beta) \pi(\beta | \tau) d \beta = \hat{R}(\tau) (1 + O_{p}(n^{-1})).
\end{equation*}
\cite{barndorff-nielsen1983} derived the magic formula to determine $f_{\tau} (\hat{\tau} | \hat{\beta})$ for the MLEs $\hat{\theta} = (\hat{\beta}, \hat{\tau})$. Under the parameter orthogonality of $\tau$ and $\beta$, \cite{cox87} showed that
\begin{equation*}
    f_{\tau} (\hat{\tau} | \hat{\beta}) = \hat{R}(\tau) (1 + O_{p}(n^{-1})).
\end{equation*}
Thus, we can view the Cox-Reid result as a case in which the conditional likelihood can be applied to eliminate nuisance fixed parameters. Note that
\begin{equation*}
     R(\tau) = f_{\tau} (\hat{\tau} | \hat{\beta}) (1 + O_{p}(n^{-1})).
\end{equation*}
Thus, we propose to call, in this paper, the integrated likelihood, namely the marginal posterior under $\pi(\beta | \tau) = 1$,
\begin{equation*}
    R(\tau) = \int L_{m}(\tau, \beta) d \beta = \int \int H(\tau, \beta, z) dz d\beta
\end{equation*}
the restricted likelihood. With the ELA, $R(\tau)$ can always be computed, as shown below, whereas $f_{\tau}(\hat{\tau} | \hat{\beta})$ is hardly available. The use of $R(\tau)$ does not require parameter orthogonality of \cite{cox87}, which would be hard to verify in general random effects models. From a frequentist perspective, the use of the integrated likelihood to eliminate the nuisance parameters has been examined for predicting unobserved latent variables $z$ by \cite{leekim2016}.

When the marginal likelihood $\ell_{m}(\theta)$ is not available, \cite{leenelder2001} proposed the use of the extended restricted likelihood
\begin{equation*}
\hat{r}(\tau )= \log \hat{R}(\tau) = h(\tau, \tilde{\beta}, \tilde{z})-\frac{1}{2}\log
\left| \frac{1}{2\pi }\tilde{\Omega}_{\psi \psi }\right| ,
\end{equation*}
where $\psi =(\beta ,z)$, $\tilde{\psi}=\OP{arg}\max_{\psi } h(\beta ,\tau ,z)$ and $\tilde{\Omega}_{\psi \psi }=\{-\partial ^{2}h(\beta ,\tau ,z)/\partial \psi \partial \psi ^{\OP{T}}\}|_{\psi =\tilde{\psi}}$. In this paper, we refer to $\hat{r}(\tau) = \log \hat{R}(\tau)$ as the approximate restricted log-likelihood. Similar to (\ref{eq:predictive_lik_LA}), the restricted likelihood $R(\tau)$ can be approximated by using the approximate predictive likelihood $\hat{L}_{p}(\psi |y;\tau )$ based on a normal distribution
\begin{equation*}
\psi \mid y \sim\OP{N}\left( \tilde{\psi},\tilde{\Omega}_{\psi \psi }^{-1}\right).
\end{equation*}
Thus, Lee and Nelder's (\citeyear{leenelder2001}) extended restricted likelihood $\hat{R}(\tau ;\psi )=H(\tau ,\psi )/\hat{L}_{p}(\psi |y;\tau )$ is the LA to $R(\tau)$. In normal linear mixed models, $R(\tau) = \hat{R}(\tau) = f_{\tau}(\hat{\tau} | \hat{\beta})$ becomes the restricted (or residual) likelihood of \cite{patterson1971}: see Chapter 5 of \cite{leenelderpawitan2017}.

We explore how to use the ELA to obtain the REMLE. Let 
\begin{equation*}
\hat{R}_{B}(\tau )=\frac{1}{B}\sum_{b=1}^{B}\hat{R}(\tau ;\psi _{b}),
\end{equation*}
where $\{\psi_{b}:b=1,\ldots ,B\}$ are iid samples from $\OP{N}(\tilde{\psi},\tilde{\Omega}_{\psi \psi }^{-1})$. Then, it is immediate that
\begin{equation*}
    \hat{r}_{B}(\tau )=\log \hat{R}_{B}(\tau )\overset{\OP{P}}{\rightarrow } r(\tau ) = \log R(\tau)
\end{equation*}
as $B \to \infty$. Moreover, let $J(\tau) = - \partial^{2} r(\tau) / \partial \tau \partial \tau^{\T}$ and $\hat{J}_{B} = J_{B}(\hat{\tau}_{B}^{\OP{ELA}})$, where
\begin{eqnarray*}
\hat{\tau}_{B}^{\OP{ELA}} &=& \OP{arg} \max_{\tau} \hat{R}_{B}(\tau), \\
    J_{B}(\tau) &=& \left [ \sum_{b=1}^{B} \left \{ \zeta(\tau, \psi_{b}) \frac{\partial h(\tau, \psi_{b})}{\partial \tau} \right \} \right ] \left [ \sum_{b=1}^{B} \left \{ \zeta(\tau, \psi_{b}) \left ( \frac{\partial h(\tau, \psi_{b})}{\partial \tau} \right )^{\T}  \right \} \right ] \nonumber \\
    && - \sum_{b=1}^{B} \left [ \zeta(\tau, \psi_{b}) \left \{ \frac{\partial h(\tau, \psi_{b})}{\partial \tau} \left ( \frac{\partial h(\tau, \psi_{b})}{\partial \tau} \right )^{\T} + \frac{\partial^{2} h(\tau, \psi_{b})}{\partial \tau \partial \tau^{\T}} \right \} \right ],
\end{eqnarray*}
and $\zeta(\tau, \psi_{b}) = \hat{R}_{m}(\tau, \psi_{b}) / \sum_{t=1}^{B} \hat{R}_{m}(\tau, \psi_{t})$. Then, we have the following theorem.
\begin{thm}
\label{thm:reml} 
Let $\hat{\tau} = \OP{arg} \max_{\tau} r(\tau)$ be the REMLE of $\tau$. As $B\rightarrow \infty $,
\begin{itemize}
    \item[] (i) $\hat{\tau}_{B}^{\OP{ELA}}\overset{\OP{P}}{\rightarrow }\hat{\tau}$,
    \item[] (ii) $\hat{J}_{B} \overset{\OP{P}}{\to} J (\hat{\tau})$. 
\end{itemize}
\end{thm}
Thus, the variance estimator of the REMLE $\hat{\tau}$ can be consistently estimated by $\widehat{\V} ( \hat{\tau} ) = \hat{J}_{B}^{-1}$. The second-order LA is applicable to only a limited class of models; for example, it cannot be applied to models with correlated random effects. The current version of the second-order LA in the dhglm in R \citep{dhglm2018} allows only crossed models with two independent random effects. However, the ELA is applicable to any statistical models with latent variables, as illustrated below.

\section{Salamander Mating Data}
\label{section:ELA_salamander} 

In this paper, we investigate how to obtain the frequentist MLE and REMLE, as well as their variance estimators. From a Bayesian perspective, the MLE and its variance estimator for $\theta = (\beta, \tau)$ are the posterior mode and its variance under a flat prior on $\theta$, whereas the REMLE and its variance estimator for $\tau$ are the marginal posterior mode and its variance under a flat conditional prior on $\beta | \tau$. Here, we investigate the performance of the MLE, REMLE, and their variance estimators, based on the ELA, through numerical studies.

\cite{mccullagh1989} presented the salamander mating data. Three experiments were conducted to collect these data: two experiments were performed with the same salamanders in the summer and fall of 1986, and the third experiment was conducted in the fall of the same year using different salamanders. The salamander data are difficult to analyse as crossed models are required for binary data with correlated random effects. The Gauss-Hermite quadrature cannot be used due to the large value of $d$. Here, we use the ELA for the analysis. We use simulation studies with $T=200$ replications to evaluate the performance of various methods based on the following quantities: (i) Est: $\bar{\theta}=\sum_{t=1}^{T} \hat{\theta}^{(t)}/T$, (ii) SE: $\sum_{t=1}^{T}\widehat{\OP{s.e.}}(\hat{\theta}^{(t)})/T$ and (iii) SD: $\{ \sum_{t=1}^{T}(\hat{\theta}^{(t)} - \bar{\theta})^{2}/(T-1)\}^{1/2}$, where $\hat{\theta}^{(t)}$ is an estimate at the $t$th replication. To evaluate the performance of the point estimation, we compare the Est and true value of the fixed parameters. The similarity between the SE and the SD indicates the performance of the variance estimation.

\subsection{Summer Data}

\cite{shun1997} and \cite{nohlee2007} investigated the data that were collected during the summer to show how the second-order LA can be applied. The authors fitted a crossed model with $d = O(n^{1/2})$. For $i=1,\ldots ,I=20$ and $j=1,\ldots, J=20$, let $y_{ij} \in \{ 0,1\} $ be the binary outcome that indicates whether mating was successful for the $i$th female and the $j$th male. Each female was paired with six males for mating, generating in 120 observations. The authors considered the following random effects model:
\begin{equation*}
\OP{logit} \OP{P}\left( y_{ij}=1\mid z_{i}^{f},z_{j}^{m}\right)
 =x_{ij}^{\OP{T}}\beta +\sigma _{f}z_{i}^{f}+\sigma _{m}z_{j}^{m},
\end{equation*}
where $z_{i}^{f}\sim \OP{N}(0,1)$ and $z_{j}^{m}\sim \OP{N}(0,1)$ are female random effects and male random effects, respectively, which are assumed to be independent of each other. The covariates $x_{ij}$ include an intercept, the main effects Trtf and Trtm, and their interaction Trtf$\cdot $Trtm, where Trtf (Trtm) = 0, 1 for Rough Butt salamanders and Whiteside salamanders, respectively. 

\begin{table}[htp!]
\caption{Simulation results for the summer data.}
\label{tab:simul_summer}
\centering
\begin{tabular}{ccccccc}
\hline Method & Intercept & Trtf & Trtm & Trtf$\times$Trtm & $\sigma_{f}$
& $\sigma_{m}$ \\ \hline
True value & 1.06 & -3.05 & -0.72 & 3.77 & 1.22 & 1.22 \\ \hline 
MQL & 0.78 & -2.36 & -0.51 & 2.87 & 0.86 & 0.88 \\
PQL & 0.85 & -2.51 & -0.57 & 3.05 & 0.94 & 0.96 \\
CPQL & 1.25 & -3.48 & -0.90 & 4.33 & 1.09 & 1.04 \\
D\&M & 1.09 & -3.15 & -0.83 & 4.04 & 1.29 & 1.32 \\
$\hat{\ell}_{m}$ & 0.93 & -2.82 & -0.60 & 3.21 & 1.04 & 1.00 \\ 
$\hat{\ell}_{m}^{s}$ & 0.98 & -2.94 & -0.63 & 3.64 & 1.19 & 1.20 \\  \hline
$\hat{r}$ & 1.15 & -3.21 & -0.79 & 3.82 & 1.26 & 1.27 \\ 
SE ($\hat{r}$) & 0.83 & 1.08  & 0.96 & 1.12 & 0.34 & 0.35 \\
SD ($\hat{r}$) & 0.97 & 1.54  & 0.92 & 1.54 & 0.61 & 0.69 \\
$\hat{r}^{s}$ & 1.05 & -3.02 & -0.69 & 3.72 & 1.23 & 1.24 \\ 
SE ($\hat{r}^{s}$) & 0.70 & 0.90  & 0.83 & 0.97 & 0.30 & 0.29 \\
SD ($\hat{r}^{s}$) & 0.62 & 0.87  & 0.66 & 0.92 & 0.48 & 0.49 \\
\hline
$\hat{r}_{2}$ & 1.11 & -3.11 & -0.84 & 3.85 & 1.11 & 1.18 \\ 
$\hat{r}_{10}$ & 0.99 & -3.09 & -0.73 & 3.78 & 1.25 & 1.25 \\ 
$\hat{r}_{50}$ & 1.07 & -3.02 & -0.72 & 3.77 & 1.21 & 1.23 \\ 
SE ($\hat{r}_{50}$) & 0.48 & 0.75  & 0.65 & 0.96 & 0.27 & 0.28 \\
SD ($\hat{r}_{50}$) & 0.51 & 0.80  & 0.57 & 0.89 & 0.38 & 0.42 \\
\hline
\end{tabular}
\end{table}

The simulation results are presented in Table \ref{tab:simul_summer}. Here $\hat{\ell}_{m}$ ($\hat{\ell}_{m}^{s}$) represents the approximate MLE and $\hat{r}$ ($\hat{r}^{s}$) represents the approximate REMLE calculated using the first-order (second-order) LA. $\hat{r}$ and $\hat{r}^{s}$ are the HL(1,1) and HL(2,2), respectively, of \cite{nohlee2007} with the approximate MLE of $\beta$ and the approximate REMLE of $\tau$ maximizing $\hat{\ell}_{m}$ ($\hat{\ell}_{m}^{s}$) and $\hat{r}$ ($\hat{r}^{s}$), respectively. The authors also examined the performance of the penalized quasi-likelihood (PQL) and marginal quasi-likelihood (MQL) methods of \cite{breslow1993} and Drum and McCullagh's (\citeyear{drum1993}) method (D\&M). Note that the PQL method has large biases in estimating the dispersion parameters \citep{lee1996, nohlee2007}. \cite{breslowlin1995} derived a correction factor for the PQL (CPQL) to remove the asymptotic bias. \cite{nohlee2007} noted that the approximate REMLE $\hat{r}^{s}$, based on the second-order LA, produced the least bias in estimating $\theta$ among the existing methods at the time. Table \ref{tab:simul_summer} shows that the REMLEs $\hat{r}$ and $\hat{r}^{s}$ perform better than the MLEs $\hat{\ell}_{m}$ and $\hat{\ell}_{m}^{s}$. $\hat{r}_{B}$ is the ELA estimation based on $B$ random samples, where the MLE of $\beta$ and the REMLE of $\tau$ maximize $\hat{\ell}_{B}$ and $\hat{r}_{B}$, respectively. $\hat{r}_{B}$ with $B \geq 10$ improves the approximate REMLE $\hat{r}$ based on the first-order LA and $\hat{r}_{50}$ improves the approximate REMLE $\hat{r}^{s}$ based on the second-order LA. The ELA is considerably easier to implement than $\hat{r}^{s}$. To evaluate the performance of variance estimators, we compare $\hat{r}$, $\hat{r}^{s}$, and $\hat{r}_{B}$. We observe that $\hat{r}$ underestimates the SD. The SE of $\hat{r}^{s}$ and $\hat{r}_{50}$ well estimate the SDs of the mean parameters; however, for $\sigma_{f}$ and $\sigma_{m}$, both $\hat{r}^{s}$ and $\hat{r}_{50}$ underestimate the SD. This underestimation of the ELA vanishes as $n$ increases, as discussed below.

\subsection{Pooled Data}

For the pooled data from the three experiments, for which $k=1,2,3$, \cite{karim1992} considered the following model:
\begin{equation*}
\OP{logit}\left\{ \OP{P}\left( y_{ijk}=1\mid z_{i}^{f},z_{j}^{m} \right) \right\} =x_{ijk}^{\OP{T}}\beta + \Sigma _{f, k}^{1/2}z_{i}^{f} + \Sigma _{m, k}^{1/2}z_{j}^{m},
\end{equation*}
where $z_{i}^{f}=( z_{i1}^{f},z_{i2}^{f},z_{i3}^{f})^{\T} \sim \OP{N}(0, I)$ and $z_{j}^{m}=( z_{j1}^{m},z_{j2}^{m},z_{j3}^{m})^{\T} \sim \OP{N}( 0,I)$ are independent, 
\begin{equation*}
\Sigma_{f}=
\begin{pmatrix}
\sigma _{f_{1}}^{2} & \rho _{f}\sigma _{f_{1}}\sigma _{f_{2}} & 0 \\ 
\rho _{f}\sigma _{f_{1}}\sigma _{f_{2}} & \sigma _{f_{2}}^{2} & 0 \\ 
0 & 0 & \sigma _{f_{2}}^{2}
\end{pmatrix}
,~\Sigma _{m}=
\begin{pmatrix}
\sigma _{m_{1}}^{2} & \rho _{m}\sigma _{m_{1}}\sigma _{m_{2}} & 0 \\ 
\rho _{m}\sigma _{m_{1}}\sigma _{m_{2}} & \sigma _{m_{2}}^{2} & 0 \\ 
0 & 0 & \sigma _{m_{2}}^{2}
\end{pmatrix}
,
\end{equation*}
and $\Sigma _{f,k}^{1/2}$ and $\Sigma _{m,k}^{1/2}$ are the $k$th rows of $\Sigma_{f}^{1/2}$ and $\Sigma_{m}^{1/2}$, respectively. Here, $\Sigma_{f, k}^{1/2} z_{i}^{f}$ and $\Sigma_{m, k}^{1/2} z_{j}^{m}$ with $k=1, 2$ represent correlated random effects. For the pooled data, an additional covariate indicating the season (0=summer and 1=fall) is included. In terms of the dispersion parameters, $\sigma_{f_{1}}^{2}$ ($\sigma _{m_{1}}^{2}$) is the variance in the summer and $\sigma_{f_{2}}^{2}$ ($\sigma _{m_{2}}^{2}$) is the variance in the fall for female (male) salamanders. Moreover, $\rho _{f}$ ($\rho _{m}$) describes the correlation resulting from the same salamander being selected in the first two experiments. The second-order LA cannot be applied since the random effects are correlated. Among frequentist methods, for correlated random effects models, the PQL of \cite{breslow1993} and $\hat{r}$ of \cite{leenelder2001} can be applied. \cite{breslow1993} applied the PQL method under the constraints $\sigma_{m_{1}}=\sigma _{m_{2}}$ and $\rho_{m}=1$. \cite{karim1992} used the Gibbs sampler to analyse the results from a Bayesian perspective.

\begin{table}[htp!]
\caption{Estimates of $\theta$ for the pooled data. The values in the parentheses are the estimated standard errors.}
\label{tab:pooled}
\centering
\addtolength{\tabcolsep}{-4pt} 
\begin{tabular}{cccccccccccc}
\hline Method & $\beta_{0}$ & $\beta_{1}$ & $\beta_{2}$ & $\beta_{3}$ & $%
\beta_{4}$ & $\sigma_{f_{1}}$ & $\sigma_{f_{2}}$ & $\rho_{f}$ & $%
\sigma_{m_{1}}$ & $\sigma_{m_{2}}$ & $\rho_{m}$ \\ 
\hline Gibbs & 
\begin{tabular}{@{}c}
1.48 \\ 
(0.64)
\end{tabular}
& 
\begin{tabular}{@{}c}
-0.62 \\ 
(0.54)
\end{tabular}
& 
\begin{tabular}{@{}c}
-3.13 \\ 
(0.62)
\end{tabular}
& 
\begin{tabular}{@{}c}
-0.76 \\ 
(0.62)
\end{tabular}
& 
\begin{tabular}{@{}c}
3.90 \\ 
(0.72)
\end{tabular}
& 1.39 & 1.17 & -0.15 & 1.12 & 1.42 & 0.96 \\ \hline
PQL & 
\begin{tabular}{@{}c}
1.18 \\ 
(0.49)
\end{tabular}
& 
\begin{tabular}{@{}c}
-0.50 \\ 
(0.41)
\end{tabular}
& 
\begin{tabular}{@{}c}
-2.43 \\ 
(0.44)
\end{tabular}
& 
\begin{tabular}{@{}c}
-0.62 \\ 
(0.46)
\end{tabular}
& 
\begin{tabular}{@{}c}
3.01 \\ 
(0.52)
\end{tabular}
& 1.04 & 0.79 & -0.15 & 0.95 & 0.95 & 1 \\ \hline
$\hat{r}$ & 
\begin{tabular}{@{}c}
1.53 \\ 
(0.58)
\end{tabular}
& 
\begin{tabular}{@{}c}
-0.63 \\ 
(0.53)
\end{tabular}
& 
\begin{tabular}{@{}c}
-3.23 \\ 
(0.56)
\end{tabular}
& 
\begin{tabular}{@{}c}
-0.79 \\ 
(0.53)
\end{tabular}
& 
\begin{tabular}{@{}c}
4.02 \\ 
(0.59)
\end{tabular}
& 
\begin{tabular}{@{}c}
1.49 \\ 
(0.38)
\end{tabular}
& 
\begin{tabular}{@{}c}
1.12 \\ 
(0.37)
\end{tabular}
& 
\begin{tabular}{@{}c}
-0.05 \\ 
(0.24)
\end{tabular}
& 
\begin{tabular}{@{}c}
0.90 \\ 
(0.43)
\end{tabular}
& 
\begin{tabular}{@{}c}
1.44 \\ 
(0.33)
\end{tabular}
& 
\begin{tabular}{@{}c}
0.72 \\ 
(0.16)
\end{tabular}
\\ \hline
$\hat{r}_{50}$ & 
\begin{tabular}{@{}c}
1.50 \\ 
(0.60)
\end{tabular}
& 
\begin{tabular}{@{}c}
-0.63 \\ 
(0.51)
\end{tabular}
& 
\begin{tabular}{@{}c}
-3.16 \\ 
(0.56)
\end{tabular}
& 
\begin{tabular}{@{}c}
-0.76 \\ 
(0.57)
\end{tabular}
& 
\begin{tabular}{@{}c}
3.90 \\ 
(0.61)
\end{tabular}
& 
\begin{tabular}{@{}c}
1.46 \\ 
(0.46)
\end{tabular}
& 
\begin{tabular}{@{}c}
1.12 \\ 
(0.31)
\end{tabular}
& 
\begin{tabular}{@{}c}
-0.13 \\ 
(0.38)
\end{tabular}
& 
\begin{tabular}{@{}c}
0.95 \\ 
(0.37)
\end{tabular}
& 
\begin{tabular}{@{}c}
1.40 \\ 
(0.34)
\end{tabular}
& 
\begin{tabular}{@{}c}
1.00 \\ 
(0.02)
\end{tabular}
\\
\hline
\end{tabular}
\end{table}

Table \ref{tab:pooled} shows the estimation results for the pooled data obtained by various methods. It is well known that the PQL has large bias in binary data. For the ELA, we set $B=50$ for the point estimation and $B=1000$ for the standard error estimation. The approximate REMLE calculated using $\hat{r}$ differs from the true REMLE calculated using the ELA $\hat{r}_{50}$ when estimating $\rho_{m}$. The Gibbs sampler uses a flat prior for the mean parameters $\beta$ and noninformative priors $\pi(\Sigma_{f}) \propto | \Sigma_{f} |^{-2}$ and $\pi(\Sigma_{m}) \propto | \Sigma_{m} |^{-2}$ for the dispersion parameters. This approach gives results similar to $\hat{r}_{50}$, which are marginal posterior modes under flat priors. For the hypotheses
\begin{equation*}
H_{0}:~\rho_{m}=1,~H_{1}:~\rho_{m}\neq 1,
\end{equation*}
the ELA gives the likelihood ratio test $2\{ \hat{\ell}_{50}( \hat{\theta} ) - \hat{\ell}_{50}( \hat{\theta}^{0} ) \} =0.1022$, where $\hat{\theta}^{0}$ is the REMLE under the null hypothesis. Thus, we cannot reject $H_{0}$. This result indicates why the estimates of $\rho_{m}$ are often close to 1 in Table \ref{tab:pooled}. Thus, we consider a submodel with a shared random effects model in which $z_{j2}^{m}=\gamma_{m}z_{j1}^{m}$ for some $\gamma_{m}$.

\begin{table}[htp!]
\caption{Simulation results for the pooled data.}
\label{tab:pooled_simul}
\centering
\begin{tabular}{@{}cccccccccccc}
\hline
\multicolumn{2}{c}{} & $\beta_{0}$ & $\beta_{1}$ & $\beta_{2}$ & $\beta_{3}$
& $\beta_{4}$ & $\sigma_{f_{1}}$ & $\sigma_{f_{2}}$ & $\rho_{f}$ & $%
\sigma_{m}$ & $\gamma_{m}$ \\ \hline
\multicolumn{2}{c}{True value} & 1.50 & -0.65 & -3.20 & -0.75 & 3.90 & 1.45
& 1.10 & -0.15 & 1.00 & 1.50 \\ \hline
$\hat{r}$ & Est & 1.69 & -0.73 & -3.48 & -0.92 & 4.28 & 1.68 & 1.31 & -0.12 & 1.24 & 1.58 \\ 
& SE  & 0.65 & 0.58  & 0.59  & 0.58  & 0.63 & 0.37 & 0.35 & 0.23  & 0.37 & 0.59 \\ 
          & SD  & 0.75 & 0.67  & 0.72  & 0.67  & 0.84 & 0.61 & 0.39 & 0.18  & 0.49 & 0.73 \\ \hline
$\hat{r}_{50}$ & Est & 1.53 & -0.75 & -3.10 & -0.72 & 3.80 & 1.50 & 1.26 & -0.14 & 1.06 & 1.55 \\ 
& SE  & 0.61 & 0.55 & 0.56 & 0.58 & 0.62 & 0.52 & 0.37 & 0.44 & 0.45 & 0.80 \\ 
         & SD  & 0.64 & 0.52 & 0.52 & 0.60 & 0.60 & 0.43 & 0.30 & 0.41 & 0.40 & 0.75 \\ 
\hline
\end{tabular}
\end{table}

Table \ref{tab:pooled_simul} shows that the estimation performance of the ELA is better than that of $\hat{r}$ for all $\theta $. In particular, $\hat{r}$ severely underestimates the standard errors. The ELA improves the point estimation and the standard error estimation. As shown in Tables \ref{tab:simul_summer} and \ref{tab:pooled_simul}, the SE obtains better estimates of the SD for the pooled data with $n=360$ than for the summer data with $n=120$. This result implies that the ELA provides consistent standard error estimators for the REMLEs.

\section{Rongelap Spatial Data}
\label{section:ELA_rongelap} 

\cite{diggle1998} presented the Rongelap data, available at the geoRglm in R \citep{geoRglm2017}, which were obtained by the Marshall Islands National Radiological Survey, to determine whether Rongelap Island is safe with respect to radionuclide contamination. The data include gamma-ray counts $y_{i}$ of radionuclide concentrations over time $t_{i}$ at the spatial location $s_{i}$ for $i=1,\ldots ,n=157$ different locations on Rongelap Island. \cite{diggle1998} considered the following Poisson random effects model: 
\begin{equation}
y_{i}\mid z\sim \OP{Poi}\left( t_{i} \lambda_{i}\right) ,~\log \lambda_{i}=\beta _{0}+\Sigma_{i}^{1/2}z,  \label{eq:model_rongelap}
\end{equation}
where $z=\left( z_{1},\ldots ,z_{n}\right)^{\OP{T}}\sim \OP{N}(0,I)$, $\Sigma_{i}^{1/2}$ is the $i$th row of $\Sigma ^{1/2}$ and the ($i, j$)th element of $\Sigma$ is
\begin{equation}
\label{eq:cov_rongelap_1}
    \Sigma_{ij} = \exp \left \{ \phi - \exp(\alpha) \| s_{i}-s_{j}\|_{2} \right \},
\end{equation}
where $\| s_{i}-s_{j}\|_{2}$ is the distance between the $i$th location and the $j$th location.

The integrated nested Laplace approximation (INLA) in R \citep{rue2009} is a widely used Bayesian procedure for fitting spatial data. Given the prior $\pi(\theta)$, the INLA approximates the posterior $\pi(\theta | y) \propto L_{m}(\theta) \pi(\theta)$ as  $\hat{\pi}(\theta | y) \propto \hat{L}_{m}(\theta) \pi(\theta)$ based on the LA. Then, the INLA uses the approximate elementwise marginal posteriors
\begin{equation}
\label{eq:inla1}
\hat{\pi}(\theta_{k} \mid y) = \int \hat{\pi}(\theta \mid y) d \theta_{-k},
\end{equation}
where $\theta_{-k} = (\theta_{1}, \ldots, \theta_{k-1}, \theta_{k+1}, \ldots )$. Instead of (\ref{eq:cov_rongelap_1}), the INLA uses the following parametrization:  
\begin{equation}
    \label{eq:cov_rongelap_2}
    \Sigma_{ij} =  \exp \left \{ - \log 2 \pi - \alpha - 2 \xi - \exp(\alpha) \| s_{i}-s_{j}\|_{2} \right \},
\end{equation}
where $\phi = - \log 2 \pi - \alpha - 2 \xi$. The covariance model (\ref{eq:cov_rongelap_1}) is referred to as an exponential covariance function, whereas model (\ref{eq:cov_rongelap_2}) is the Mat\'{e}rn covariance function, which is adopted by the INLA \citep{moraga2019}. Under Gaussian priors for $\beta_{0}$, $\xi$, and $\alpha$, the INLA provides the mean, mode, and standard deviations using random samples from the marginal posterior (\ref{eq:inla1}).

Although the responses are counts and thus not binary, since $d = n$, the LA may not be suitable. In addition, the second-order LA cannot be used due to the correlated random effects. We fitted the original Poisson random effects model (\ref{eq:model_rongelap}), but it showed a severe lack-of-fit, with a scaled deviance of 6.466 for 0.717 degrees of freedom. If there is no lack-of-fit, the scaled deviance follows the chi-squared distribution with computed degrees of freedom. \cite{Bivand2015} proposed the overdispersed Poisson model for $y_{i} | z$:
\begin{equation}
\label{eq:rongelap_quasi}
    c_{i} \mid z \sim \OP{Poi} (\lambda_{i}), ~ \log \lambda_{i}=\beta_{0} + \Sigma_{i}^{1/2}z,
\end{equation}
where $c_{i} = y_{i} / t_{i}$. The authors fitted the model (\ref{eq:rongelap_quasi}) by using the INLA. Note that under the model (\ref{eq:rongelap_quasi}), we have an overdispersed Poisson random effects model with $\E(y_{i} | z) = t_{i} \lambda_{i} = \mu_{i}$, $\V(y_{i} | z) = t_{i}^{2} \lambda_{i} = t_{i} \mu_{i} > \mu_{i}$ and overdispersion parameters $t_{i} > 1$. \cite{leenelderpawitan2017} showed that the use of the model (\ref{eq:rongelap_quasi}) is equivalent to the use of the extended quasi-likelihood \citep{leenelder2000} for fitting an overdispersed Poisson model with $y_{i} | z$. The overdispersed Poisson model (\ref{eq:rongelap_quasi}) has a scaled deviance of 120.1 with 146.9 degrees of freedom, confirming no lack-of-fit. Thus, the overdispersed Poisson model (\ref{eq:rongelap_quasi}) achieves a better fit than the original Poisson model (\ref{eq:model_rongelap}). 

\begin{table}[htp!]
\caption{Estimates of the parameters according to the Rongelap data under the model (11). The values in parentheses are the estimated standard errors.}
\label{tab:rongelap}
\centering
\begin{tabular}{ccccc}
\hline Method & $\beta_{0}$ & $\phi$ & $\alpha$ & $\xi$ \\  \hline
$\hat{r}$ & 1.966 (0.129) & -3.051 (0.355) & -2.708 (0.827) & 1.961 (0.203) \\ \hline
$\hat{r}_{B}$ & 1.983 (0.102) & -3.325 (0.932) & -2.489 (1.424) & 1.988 (0.724) \\ \hline
$INLA$ & 2.005 (0.116) & $\cdot$  & -1.822 (0.722) & 1.886 (0.524) \\ \hline
$INLA^{\ast}$ & 1.990 (0.436) & $\cdot$  & -1.674 (0.722) & 1.770 (0.524)  \\ \hline
\end{tabular}
\end{table}

For the ELA, $B=200$ is selected to fit $\beta_{0}$, $B=1000$ is selected to fit $\tau$ and $B=2000$ is selected to estimate the standard error. The estimation results of the Rongelap data with model (\ref{eq:rongelap_quasi}) are presented in Table \ref{tab:rongelap}. For the point estimates, we consider both the posterior mean (\textit{INLA}) and posterior mode (\textit{INLA}$^{\ast}$) of the INLA output. The INLA provides a posterior standard deviation (PSD) for samples from the marginal posterior distribution as a standard error estimation. Since the Bayesian approach is not invariant with respect to the transformation of parameters, we do not report on $\phi$ for the INLA. However, ML estimation is invariant with respect to transformation; thus, we present the ELA result of $\xi$ obtained by using the delta method. The REMLEs calculated by the ELA are marginal posterior modes under flat priors; thus, the difference between the ELA and the INLA would be caused by the use of different priors, although these differences are not significant.

\begin{table}[htp!]
\caption{Simulation results for the Rongelap data.}
\label{tab:rongelap_simul}
\centering
\begin{tabular}{@{}cccccc}
\hline
\multicolumn{2}{c}{} & $\beta_{0}$ & $\phi$ & $\alpha$ & $\xi$ \\ \hline
\multicolumn{2}{c}{True value} & 1.980 & -3.000 & 0.100 & 0.531 \\ \hline
$\hat{r}$     & Est & 1.976 & -3.023  & 0.178 & 0.504 \\ 
              & SE  & 0.050 & 0.341  & 0.534 & 0.318 \\ 
              & SD  & 0.048 & 0.416  & 0.688 & 0.430 \\ \hline
$\hat{r}_{B}$ & Est & 1.977 & -3.014 & 0.119 & 0.528 \\ 
              & SE  & 0.051 & 0.476  & 0.740 & 0.442 \\  
              & SD  & 0.049 & 0.437  & 0.728 & 0.444 \\ \hline
$INLA$   & Est & 1.986 & $\cdot$  & 0.051 & 0.673 \\ 
              & PSD  & 0.087 & $\cdot$  & 0.681 & 0.602 \\  
              & SD  & 0.051 & $\cdot$  & 0.675 & 0.598 \\ \hline
$INLA^{\ast}$   & Est & 1.988 & $\cdot$  & 0.037 & 0.632 \\ 
              & PSD  & 0.087 & $\cdot$  & 0.681 & 0.602 \\ 
              & SD  & 0.051 & $\cdot$ & 0.627 & 0.595 \\ \hline
\end{tabular}
\end{table}

We perform a simulation study with model (\ref{eq:rongelap_quasi}). To reduce the complexity of using the extended quasi-likelihood method, we use a Poisson random effects model by setting $t_{i} = 1$. According to Table \ref{tab:rongelap_simul}, the point estimates of $\beta_{0}$ are similar for all the evaluated methods. In terms of the standard error estimates, the LA $\hat{r}$ underestimates the SD of the estimators. The ELA provides accurate REMLEs. We report the INLA results to highlight the differences caused by the use of different priors. The INLA computes the PSDs using samples from the marginal posteriors, whereas the standard error estimates of the REMLEs are computed using the Hessian matrix without resampling. In summary, different priors could yield different dispersion parameter estimates.

\section{Concluding Remarks}
\label{section:ELA_conclusion} 

The LA and the variational Bayes method have been proposed as methods for approximating the marginal likelihood. However, resulting approximate MLEs and REMLEs could be often biased for binary or spatial data. Furthermore, a consistent variance estimation method is not available. With the ELA, the MLE, REMLE, and their consistent variance estimators can be obtained in general for statistical models with unobserved latent variables. The results of numerical studies confirm that the ELA provides satisfactory MLE and REMLE for a wide variety of models. Furthermore, the MLE and REMLE are Bayesian posterior modes and marginal posterior modes, respectively, under flat priors. Thus, we can have both frequentist and Bayesian interpretations from ML and REML analyses.

\bibliographystyle{apalike}
\bibliography{sample.bib}

\section*{Appendix: Proofs}

\subsection*{Proof of (\ref{eq:normal_equiv})}

Suppose that the true predictive likelihood $L_{p}(z|y; \theta)$ is
from a normal distribution. Let $m$ and $S$ be mean and covariance matrix of normal distribution of which predictive log-likelihood is 
\begin{equation*}
\ell_{p} \left ( z \mid y; \theta \right ) = - \frac{1}{2} \log \left | 2
\pi S \right | - \frac{1}{2} (z - m)^{\OP{T}} S^{-1} (z - m).
\end{equation*}
Then, $\tilde{z} = m$ and $\tilde{\Omega}_{zz} = S^{-1}$ since 
\begin{eqnarray*}
\frac{\partial}{\partial z} \ell_{p} \left ( z \mid y; \theta \right ) &=& - S^{-1} (z - m), \\
\frac{\partial}{\partial z \partial z^{\OP{T}}} \ell_{p} \left ( z \mid y;
\theta \right ) &=& - S^{-1}.
\end{eqnarray*}
Thus, we have $\hat{L}_{p}(z|y; \theta) = L_{p}(z|y; \theta)$ for all $z$ which gives $\hat{L}_{m}(\theta) = L_{m}(\theta)$. Moreover, 
\begin{equation*}
\hat{L}_{B}(\theta) = \frac{1}{B} \sum_{b=1}^{B} \frac{H(\theta, Z_{b})}{\hat{L}_{p}(Z_{b} \mid y; \theta)} = \frac{1}{B} \sum_{b=1}^{B} \frac{H(\theta, Z_{b})}{L_{p}(Z_{b} \mid y; \theta)} = L_{m}(\theta),
\end{equation*}
for all $B \geq 1$.

\subsection*{Proof of Theorem \ref{thm:ELA}}

Note that there exists a constant $M > 0$ such that 
\begin{equation}
\label{eq:thm_bounded}
\hat{L}_{m}(\theta; Z) = \frac{H(\theta, Z)}{\hat{L}_{p}(Z \mid y; \theta)}
\leq \frac{H(\theta, \tilde{z})}{\hat{L}_{p}(Z \mid y; \theta)} \leq M
\end{equation}
with probability one, i.e., $\hat{L}_{m}(\theta; Z)$ is bounded with probability
one. By the law of large numbers, we have 
\begin{equation*}
\hat{L}_{B}(\theta) = \frac{1}{B} \sum_{b=1}^{B} \frac{H(\theta, Z_{b})}{\hat{L}_{p}(Z_{b} \mid y; \theta)} \overset{\OP{P}}{\to} \int \frac{
H(\theta, z)}{\hat{L}_{p}(z \mid y; \theta)} \hat{L}_{p}(z \mid y;
\theta) dz = L_{m}(\theta)
\end{equation*}
as $B \to \infty$ for all $\theta$. Then, from the Theorem 2.7 of \cite{newey1994}, we can conclude that $\hat{\theta}_{B}^{\OP{ELA}} \overset{\OP{P}}{\to} \theta$.

\subsection*{Proof of Theorem \ref{thm:variance}}
Note that the Hessian matrix of the marginal log-likelihood can be expressed as
\begin{equation}
\label{eq:Hessian_ELA}
    \frac{\partial^{2} \ell_{m}(\theta)}{\partial \theta \partial \theta^{\T}} = - \left \{ \frac{1}{L_{m} (\theta)} \frac{\partial L_{m} (\theta)}{ \partial \theta} \right \} \left \{ \frac{1}{L_{m}(\theta)} \left ( \frac{\partial L_{m}(\theta)}{ \partial \theta} \right )^{\T} \right \} + \frac{1}{L_{m}(\theta)} \frac{\partial^{2} L_{m}(\theta)}{\partial \theta \partial \theta^{\T}}.
\end{equation}
By introducing an arbitrary density function $q(z)$, we have
\begin{eqnarray*}
\frac{\partial L_{m}(\theta)}{\partial \theta} &=& \int \frac{\partial h(\theta, z)}{\partial \theta} \frac{H(\theta, z)}{q(z)} q(z) dz, \\
\frac{\partial^{2} L_{m}(\theta)}{\partial \theta \partial \theta^{\T}} &=& \int \left \{ \frac{\partial h(\theta, z)}{\partial \theta} \left ( \frac{\partial h(\theta, z)}{\partial \theta} \right )^{\T} + \frac{\partial^{2} h(\theta, z)}{\partial \theta \partial \theta^{\T}} \right \} \frac{H(\theta, z)}{q(z)} q(z) dz.
\end{eqnarray*}
Recall that
\begin{equation}
\label{eq:gradient_ELA}
    \frac{\partial \ell_{m}(\theta)}{\partial \theta} = \frac{1}{L_{m}(\theta)} \frac{\partial L_{m}(\theta)}{\partial \theta} = \frac{1}{L_{m}(\theta)} \int \frac{\partial h(\theta, z)}{\partial \theta} \hat{L}_{m}(\theta, z) \hat{L}_{p}(z \mid y; \theta) dz
\end{equation}
and $\hat{L}_{B}(\theta) \overset{\OP{P}}{\to} L_{m}(\theta)$ as $B \to \infty$. By assumption of unimodality, there exists $\{ \hat{\theta}_{B}^{\OP{ELA}}, \hat{\theta} \} \in \Theta_{1} \subset \Theta$ such that
\begin{equation*}
     \sup_{\theta \in \Theta_{1}} \left | \frac{\partial h(\theta, z)}{\partial \theta} \right | \leq M_{1}, ~ \sup_{\theta \in \Theta_{1}} \left | \frac{\partial^{2} h(\theta, z)}{\partial \theta \partial \theta^{\T}} \right | \leq M_{2}
\end{equation*}
given $M_{1}, M_{2} > 0$ for all $z$. Moreover, $w(\theta, Z)$ is bounded provided by (\ref{eq:thm_bounded}). Then,
\begin{equation}
\label{eq:proof_score}
    \frac{1}{B} \sum_{b=1}^{B} \frac{\partial h(\theta, Z_{b})}{\partial \theta} \hat{L}_{m}(\theta, Z_{b}) \overset{\OP{P}}{\to} \frac{\partial L_{m}(\theta)}{\partial \theta}.
\end{equation}
By using the Slutsky's theorem, we have
\begin{equation*}
    \frac{\frac{1}{B} \sum_{b=1}^{B} \frac{\partial h(\theta, Z_{b})}{\partial \theta} \hat{L}_{m}(\theta, Z_{b})}{\frac{1}{B} \sum_{t=1}^{B} \hat{L}_{m}(\theta, Z_{t})} = \sum_{b=1}^{B} \frac{\partial h(\theta, Z_{b})}{\partial \theta} w(\theta, Z_{b}) \overset{\OP{P}}{\to} \frac{\partial \ell_{m}(\theta)}{\partial \theta}
\end{equation*}
as $B \to \infty$. Similar to (\ref{eq:proof_score}), we also have
\begin{equation*}
    \frac{1}{B} \sum_{b=1}^{B} \left \{ \frac{\partial h(\theta, Z_{b})}{\partial \theta} \left ( \frac{\partial h(\theta, Z_{b})}{\partial \theta} \right )^{\T} + \frac{\partial^{2} h(\theta, Z_{b})}{\partial \theta \partial \theta^{\T}} \right \} \hat{L}_{m}(\theta, Z_{b}) \overset{\OP{P}}{\to} \frac{\partial^{2} L_{m}(\theta)}{\partial \theta \partial \theta^{\T}}
\end{equation*}
which implies
\begin{equation}
\label{eq:proof_Hessian}
    \sum_{b=1}^{B} \left \{ \frac{\partial h(\theta, Z_{b})}{\partial \theta} \left ( \frac{\partial h(\theta, Z_{b})}{\partial \theta} \right )^{\T} + \frac{\partial^{2} h(\theta, Z_{b})}{\partial \theta \partial \theta^{\T}} \right \} w(\theta, Z_{b}) \overset{\OP{P}}{\to} \frac{1}{L_{m}(\theta)} \frac{\partial^{2} L_{m}(\theta)}{\partial \theta \partial \theta^{\T}}.
\end{equation}
By combining (\ref{eq:proof_score}) and (\ref{eq:proof_Hessian}), we have
\begin{equation*}
    I_{B}(\theta) \overset{\OP{P}}{\to} I(\theta) = - \frac{\partial^{2} \ell_{m}(\theta)}{\partial \theta \partial \theta^{\T}}
\end{equation*}
as $B \to \infty$ for $\theta \in \Theta_{1}$. By definition, $\Theta_{1}$ contains $\hat{\theta}_{B}^{\OP{ELA}}$ and $\hat{\theta}$. Also, $\hat{\theta}_{B}^{\OP{ELA}}$ converges to $\hat{\theta}$ as shown in Theorem \ref{thm:ELA}. In conclusion, $\hat{I}_{B}$ converges to $\hat{I}$ as $B \to \infty$ which proves the Theorem \ref{thm:variance}.
\end{document}